\RequirePackage{amsmath}
\documentclass[runningheads]{llncs}
\usepackage[T1]{fontenc}
\usepackage{xcolor}
\usepackage{graphicx}
\usepackage[noend, ruled,lined,linesnumbered]{algorithm2e}
\usepackage{float}
\begin{document}
\title{Social Deliberation vs.\ Social Contracts in
Self-Governing Voluntary Organisations}
\titlerunning{Social Deliberation vs.\ Social Contracts}
%
\author{Matthew Scott\inst{1} \and
Asimina Mertzani\inst{1} \and
Ciske Smit\inst{1} \and \\
Stefan Sarkadi\inst{2} \and
Jeremy Pitt\inst{1}}
\authorrunning{Scott et al.}
%
\institute{Department of Electrical \& Electronic Engineering, Imperial College London \\
\and
Department of Informatics, King's College London
\email{}}
\maketitle              
\begin{abstract}
Self-organising multi-agent systems regulate their components' behaviour
voluntarily, according to a set of socially-constructed, mutually-agreed, and mutable
\textit{social arrangements}.
In some systems, these arrangements may be applied with a frequency, at a scale and within
implicit cost constraints such that performance becomes a pressing issue.
This paper introduces the \textit{Megabike Scenario}, which consists of a negotiated agreement
on a relatively `large' set of conventional rules, `frequent' `democratic' decision-making
according to those rules, and a resource-bounded imperative to reach `correct' decisions.
A formalism is defined for effective rule representation
and processing in the scenario, and is evaluated against five
interleaved \textit{socio-functional} requirements. 
System performance is also evaluated empirically through simulation.
We conclude that to self-organise their social arrangements, agents need some
awareness of their own limitations and the value of compromise.

\end{abstract}

\section{Introduction} \label{sec:intro}

In a self-organising multi-agent system \cite{Bible21},
the component agents voluntarily agree to regulate
their own behaviour according to a set of socially-constructed,
mutually-agreed, and mutable \textit{social arrangements}.
Informally introduced in \cite{GraeberWengrow}, we define ``social arrangements''
as an umbrella term for any type of conventional rule-based system that
members of a group agree on for voluntarily regulating behaviour and holding themselves
accountable to one another, whether this is a convention, norm, procedure, regulation,
institution, contract or law \cite{Ostrom90,RONR00,SE11}.

In principle, these social arrangements support self-governance through deliberative processes, whereby those who are affected by the arrangements participate in their selection, modification and
enforcement, and self-determine certain configurations of the social arrangement to be congruent with, or fit-for-purpose for, prevailing environmental conditions \cite{Ostrom90,GraeberWengrow}. 
In practice, the social arrangements may need to be applied with a frequency, at a scale, and within
implicit cost constraints such that performance becomes a pressing issue, and especially so in the
presence of existential threats.

In such circumstances, it may be necessary to circumvent resource-intensive deliberative processes,
not by reducing participation through sortition or ``representative democracy'', but by
using instead \textit{social contracts}, which combine an expressive rule representation combined with efficient rule processing.
However, given the centrality of `correctness' in the outcomes of deliberative decision-making\footnote{For example, Ober \cite{jvp:ref:Obe08} attributes the success of the classical Athenian democracy
in outperforming other city-states, despite parity in other metrics, to its superior knowledge
management processes which produced `better' decisions more often.}, this demands that an alternative approach based on social contracts needs to be evaluated
against five interleaved \textit{socio-functional requirements}, which concern the entanglement
of social arrangements with computational operation, namely: 
\begin{itemize}
    \item \textit{Scalability}: do the social arrangements scale with the number of agents, the number of
rules, the cost and frequency of applying the rules, etc.?
\item \textit{Complexity}: are the social arrangements congruent with the `cognitive' ability of the agents
to use them, and do they correspond to the difficulty of the problem to be addressed
(cf.\ \cite{Rych21})?
\item \textit{Mutability}: how `easy' or `quick' is it to change both the social arrangements, 
and agreements underpinning the choice of social arrangement, especially in 
real-time, existential-risk situations (cf.\ \cite{SP23})?
\item \textit{Enforceability}: can sanctions specified by the rules be  enforced effectively
in decentralised systems with no form of coercion? and
\item \textit{Versatility}: how `seamlessly' can the social arrangements, and
conceptual resources that are socially-constructed externalities produced by
applying these arrangements, be transposed to another context?
\end{itemize}


Therefore, the motivating problem for this work is the reduction in the burden of self-governance 
by the use of expressive and tractable \text{social contracts} \cite{PREVCOINE}
instead of expensive and potentially intractable deliberative processes.
These social contracts are evaluated against the socio-functional requirements 
in the context of the \textit{Megabike
Scenario}. In this scenario, autonomous agents have to, first, self-select membership
of a group; secondly, within the group, negotiate an agreement
on a relatively `large' set of social arrangements; and thirdly make `frequent' `democratic' decisions
according to these arrangements, within a resource-bounded imperative to reach `correct' decisions.
Being used to resolve iteratively several inter-dependent social dilemmas (including
scarce resource allocation and collective risk), the social arrangements are subject to
the identified performance issues, and so social contracts preferred to deliberative
processes.

Accordingly, this paper is structured as follows. The next section introduces the 
\textit{Megabike Scenario}, while Section~\ref{sec:rules} summarises the `decision-making arenas'
 required by the scenario and exposes the limitations of social deliberation. Section~\ref{sec:formalism} specifies a formalism for effective representation
 of social contracts, as an alternative to social deliberation,
after which Section~\ref{sec:complexity} evaluates the formalism
against the socio-functional requirements. 
The performance improvements are evaluated empirically through a simulation 
of the \textit{Megabike}, described in Section~\ref{sec:simulation}.
After a discussion of related research in Section~\ref{sec:furtherWork},
we conclude in Section~\ref{sec:conclusions} that the significance of these 
results for self-organising multi-agent systems is that the agents need an acute
self-awareness of their own limitations in selecting and modifying their social arrangements.


\section{The \textit{Megabike} Scenario}\label{sec:megabike}

A \textit{megabike}, based on ``real world'' bikes for multiple riders, allows
a group of otherwise autonomous agents to occupy a single vehicle collectively,
and then propel it (pedal) to navigate a typical AI/multi-agent gridworld, both in
search of rewards (\textit{lootboxes}) and to avoid an existential threat.
Each agent is individually capable of pedalling, braking, and steering the \textit{megabike};
 consequently, the agents must collectively
 agree on (and each agent explicitly agrees to voluntarily
comply with) the social arrangements that determine direction (steerage), effort (pedalling
and braking), lootbox targeting and loot allocation, and assignment of social roles.

Ultimately each agent wants to maximise its own utility,  measured in terms of 
energy gained from lootboxes, duration of survival, and/or others' appreciation of its individual
contribution to a collective effort (e.g.,\ energy expended, compromising, performance of a social role, etc.). 
The scenario therefore involves:
\begin{itemize}
 \item \textit{institutional foundation}, as the agents must negotiate firstly,
 with whom to team up with to occupy
 a \textit{megabike}, and secondly, what social arrangements (form of voluntary
 self-governance) are agreed to be in force;
 \item \textit{coordination}, in terms of selecting a target lootbox and investing personal resources (energy)
 into a collective effort to move there; and avoiding both contradictory actions (pedalling and
 braking at the same time) or duplicating actions (causing e.g., over-steering);
 \item \textit{distributive justice}, in terms of how to allocate rewards from the successful appropriation
 of the contents of a lootbox, according to negotiated criteria also specified by the social arrangements; 
 \item \textit{normative compliance}, in terms of compliance, or otherwise, with group decisions made
 according to the agreed social arrangements, and punishment for non-compliance (although
 the collective with regards to$\ldots$)
 \item $\ldots$\textit{cooperative survival}, in that the \textit{megabike} needs sufficient occupants with sufficient energy in order to increase its chances of acquiring lootboxes before other \textit{megabike}s take them,
 and avoiding an existential threat;
 \item \textit{contributive justice}, in the form of opportunities to contribute meaningfully to successful
 collective action, norm compliance, compromises, etc., and to be appreciated for such contributions;
 and
 \item \textit{social construction}, in the form of conceptual resources like esteem, 
 trustworthiness, social networks, etc., which
 start from zero but accumulate (are socially-constructed) over time.
 \end{itemize}

An informal specification of the \textit{Megabike} Scenario `game loop' is given in 
Algorithm~\ref{alg:gameloop}. The `game' is played over multiple iterations, each iteration
consisting of multiple rounds. At the start of each iteration, the agents negotiate membership
of, and social arrangements for, their own \textit{megabike}, and perform role assignment
(in particular identifying a \textit{leader}). In each round, they select a target lootbox,
then commit some of their own energy to pedalling, braking or steering
actions. If they succeed in acquiring a lootbox, then they have to distribute the resources
according to the negotiated social arrangements. Finally, they follow protocols to exclude
an agent (ultimate sanction for non-compliance) or admit a new agent (if there is an unfilled `seat' on the \textit{megabike}).
\begin{algorithm}
$i \leftarrow 0$ ; initialise agents \; 
\Repeat{$i == $ MaxIterations OR deadlock}{
	(within agents) negotiate \textit{megabike} membership \tcp*{\hspace*{-0.58em}self-selection phase}
	(within \textit{megabike}) negotiate social arrangements \tcp*{\hspace*{-0.58em}action phase}
	(within \textit{megabike}) perform role assignment \;
	$j \leftarrow 0$ \;
	\Repeat(\CommentSty{$\quad\quad$ // operation phase}){$j == $ MaxRounds OR \textit{megabike} terminated}{
		(within \textit{megabike}) decide target lootbox \;
		(within \textit{megabike}) each agent decides pedal/brake/steer actions\;
		(environment) apply effects of actions \;
		\If{lootbox reached}{
			(within \textit{megabike}) apply resource allocation
		}
		
		(within \textit{megabike}) admission/exclusion \;
		\textbf{inc}($j$) \;
	}
	\textbf{inc}($i$) \;
}
\caption{\textit{Megabike} Scenario `Game Loop'}\label{alg:gameloop}
\end{algorithm}

Therefore, a \textit{Megabike} game (in the simulation sense) consists of several (inter-related) sub-games (in the game-theoretic sense), including:
\begin{itemize}
\item a veil of (decreasing) ignorance dilemma \cite{Rawls71}: at the start of each iteration, knowledge about
negotiating which bike
to join and which social arrangements to adopt changes over time, from a ``risk'' trust decision
to a ``reliance'' trust decision as information is gained about the behaviour of other agents;
\item a pair of collective action dilemmas \cite{Ostrom90}:
the optimal utility-maximisation and self-preservation
strategy is (literally) free-riding, by not  
expending energy while other agents do all the pedalling; but if all agents use this strategy,
all suffer, because being stationary is unsustainable as the \textit{megabike} will be terminated by
the existential threat or competition with other
\textit{megabikes} for lootboxes;
\item a resource allocation dilemma \cite{Rescher66}, where the allocation of scarce resources
gathered from any lootboxes must be decided 
according to a protocol included in the set of agreed social arrangements
(rather than a brutal  `real time' grab); and
\item a balloon debate dilemma, in which some agent might have to convince the others that it should
not be excluded, or that another agent should be excluded (e.g.,\ for reasons of non-compliance
with the agreed social arrangements or leadership decisions).
\end{itemize}


As per \cite{Ostrom90} and \cite{Ober17}, resolving these dilemmas
requires the negotiation of social arrangements, as discussed in the next
section.

\section{Social Deliberation for Self-Governing \textit{Megabikes}}\label{sec:rules}

This section considers decision arenas in which social arrangements
are required to  negotiate agreements and reach decisions through
processes of social deliberation.
However, it also exposes some of the situational limitations of unrestricted
social deliberation.


\subsection{Social Deliberation} \label{sec:deliberation}

As indicated by Algorithm~\ref{alg:gameloop}, the individual demands on agents in the
\textit{Megabike} Scenario are manifold, but also they need to engage in (potentially)
substantial processes of action selection, 
collective (social) deliberation and decision-making. For the former, 
each agent will need its individual strategy, which is not discussed further here, except to the 
extent that preferences inform the
latter. For this, deliberation and decision-making, they need a relatively `large' set of rules or procedures (i.e., their social arrangements).

These social arrangements are negotiated once a group of agents have all agreed to occupy a megabike,
often involving a particular type of rule, and then parameters for that rule.
For example,  decisions might be made by the assigned \textit{leader}, or by 
majority vote; if they take a vote, they have to decide which of many voting methods
to use (e.g., plurality, alternative vote, Borda count, etc.).

Rules are derived from multiple sources, including self-governing institutions
for common-pool resource management \cite{Ostrom90}, norm-governed multi-agent
systems using institutionalised power \cite{ArtikisXY}, distributive justice using
legitimate claims \cite{Rescher66}, the Game of Nomic \cite{Suber90}, Robert's Rules
of Order \cite{RONR00}, Basic Democracy \cite{Ober17} and democracy-by-design
\cite{PO18}. As such, the decision
arenas \cite{Ostrom90} include:

\textit{Mutability} \cite{Suber90,Ober17}: an agreement is required on whether or not
to distinguish between mutable and immutable rules. If a distinction is agreed, then there
needs to be one protocol for converting a mutable rule to an immutable rule, and another
protocol for converting an immutable rule to a mutable one.

\textit{Role Assignment} \cite{ArtikisXY}: minimally, a \textit{leader} will be assigned, who
will then have the institutionalised power to assert certain (institutional) facts \cite{JS96}.
A protocol for selection is required, alternative methods include by vote, weighted criteria,
torno (if equal participation is a key principle \cite{Ober17}), and historical (auto-autocrat).
Equally, a protocol for de-selection is required, if the elected representative takes undue
advantage of its position \cite{PO18}.
Note, the collective might delegate other tasks to `responsible' agents to minimise costs
(e.g.,\ steering, observing, etc.), which opens up further opportunities for investigating deception.

\textit{Lootboxes} \cite{Ostrom90,Rescher66}: a protocol for lootbox target selection is
required: this might entail a phase of `democratic' deliberation (knowledge aggregation)
with respect to social welfare optimisation, and a method for knowledge alignment, 
i.e.,\ having taken the decision, how to ensure that all agents act upon it effectively.
The \textit{leader} might, for instance, use its institutionalised power to oblige all
agents to pedal with a certain intensity. Having acquired a lootbox, a protocol for
resource allocation is required. If there is no protocol fixed, the agents might try
to `grab' loot, i.e.,\ this introduces a hawk-dove game to the scenario. 

\textit{Membership} \cite{Ostrom90}: a key feature of collective action decision arenas are
the boundaries on who is and is not constrained by the social arrangements, and which are
and are not common interest (rather than factional) issues. However, since the social
arrangements are conventional, non-compliance is a possibility; if the ultimate sanction
is exclusion then a protocol for this is required (implying pre-conditions, process, appeals
procedure, etc. \cite{Bible21}). Similarly, and since agents may be eliminated due to
depleted energy, an admissions protocol is also required: since there are various alternative
admissions processes, this is another parameter that needs to be negotiated.

\textit{Monitoring and Sanctions} \cite{Ostrom90}: assuming an open system, there is no full disclosure,
and agents cannot see how much effort other agents are putting into pedalling. However,
they can know if someone under-contributed, and for this reason, there needs to be an
auditing protocol. This would require another role assignment, a choice between methods
and outcomes (e.g.,\ cheap but unlikely to reveal non-compliance, expensive but likely).
Note that a `monetised' obligation for monitoring and auditing also exposes risks of
non-compliance with the role (doing the work, reporting results honestly, etc.). Therefore, a graduated scale of sanctions for non-compliance
needs to be negotiated.


\textit{Crisis Response}: to deal with an existential threat, a switch between decentralised
and centralised decision-making might be negotiated. The problem is to ensure that
once the crisis has passed (if it existed in the first place \cite{ShockDoctrine}), democratic
backsliding is averted and the social arrangements revert to their original form, and
do not get stuck in an autocratic or hierarchical regime \cite{GraeberWengrow}.

\subsection{The Limitations of Social Deliberation}

In the \textit{Megabike} scenario, 
given the number of decision arenas, the frequency of decision-making
in those arenas, the cognitive and communicative overheads of social
deliberation, and the pressure of an existential threat, there are
limits on the use deliberative processes, and a more efficient but equally effective way of reaching a `correct'
decision is required.

There are, in fact, precedents for substituting complex procedures 
in self-organising multi-agent systems.
For example, in previous work, e.g., \cite{ArtikisXY}, 
the focus of attention was on
events which determined the (institutionalised) powers, permissions
and obligations of agents. For this purpose, the Event Calculus (EC)
was an effective tool for executable specification, but `simple' Prolog
implementations of the EC proved unsuitable for simulations with large numbers
of agents or `long' narratives.
For experimental simulation,
EC-based logical representation of Ostrom's design principles
for self-governing institutions \cite{Ostrom90}
were re-implemented in procedural or object-oriented
languages to address issues of scale and run-time
(\cite{PSA12,PBM14}).
For real-world applications, this limitation of the EC
led to the development of a logic formalism for large-scale
run-time event recognition \cite{ArtikisRTEC}. 

Fuirthermore, in the implementation of the thought experiment Demopolis \cite{Ober17},
the rules were
conceived as defining a specification space of $n$ rules each with
$m$ parameters, with each parameter having $x$ values, so 
the `$k$-th' rule, $R_k$, was
defined as in Equation~\ref{equ:ruleRep}. 
\begin{equation}\label{equ:ruleRep}
    R_k = \begin{bmatrix}
                        P_{1,k}(V_{1_{1,k}},\ldots,V_{x_{1,k}}) \\
                        P_{2,k}(V_{1_{2,k}},\ldots,V_{x_{2,k}}) \\
                        \vdots \\
                        P_{m,k}(V_{1_{m,k}},\ldots,V_{x_{m,k}})
                      \end{bmatrix}
\end{equation}
While this meant that rules could be conveniently
written and executed as Prolog predicates, changing 
parameters meant retracting and re-asserting clauses,
and changing sets of rules meant re-consulting files.
Both of these operations are relatively `slow',
and so mutability was correspondingly problematic \cite{PO18}.

To compound matters, for agents embedded in the actual simulation,
the processes of social deliberation, 
as previously enumerated, incur substantial overheads
in communication, especially as the number of agents, and the
frequency with which deliberation occurs, scale upwards. 
These overheads cause obvious problems in situations where
cooperative survival is a condition of continued participation,
both individually and collectively (i.e.,\ no-one survives
unless everyone survives).
There are further problems of asynchrony and concurrency
that open distributed systems have to address, especially issues
of timing, sequence and causality in systems without global clocks,
which generally
do not present comparable difficulties for social systems.

In the context of the \textit{Megabike Scenario},
we propose that during the rule selection phase, the
agents should also determine whether or not deliberation can be
replaced instead by a \textit{social contract} \cite{PREVCOINE}.
This is not going against a principle of Democracy-by-Design
\cite{PO18}, ``no short-cutting democratic processes'', but instead
is seeking to define a more efficient, and mutually-agreed, 
rule-based alternative to 
social deliberation. Note that this proposal implies that in the (second) entrenchment phase of negotiating the social arrangements
for a \textit{megabike}, the $\Theta$-Learning algorithm \cite{MOP23}
could be used for reaching consensus while exploiting compromise and
dissent as conceptual resources; and in the (third) operationalisation
phase, there is an opportunity to use learning algorithms to 
customise the social contract by establishing the pathways
to requisite social influence \cite{ICAART}.

Therefore, we need an appropriate (i.e.,\ computationally tractable)
rule representation which can be used as a surrogate for 
(potentially computationally intractable) social deliberation:
for example, \textit{social contracts}.

\section{Social Contracts for Self-Governing \textit{Megabikes}}
\label{sec:formalism}

In this section, we specify \textit{social contracts} as a set of rules,
which effectively prune the search space of possible decisions.
In this way, we can reframe social deliberation as a social contract, whereby rule application and mutability are re-interpreted as optimisation problems.

In this case, agents can avoid deliberation by mutually agreeing on a set of rules that offsets the deliberation process by approximating it instead. For example, rules may be imposed using quantitative features of the lootboxes (such as relative distance or reward) to restrict the possible set of lootboxes that may be voted on by the agents. A `well-optimised' ruleset would be one that effectively removes the need for social deliberation, as after pruning the full set of lootboxes, only one is left. This process, while not deliberative, yields the same outcome: a single decision resulting from an initial search space of multiple valid decisions. As such, the ruleset serves as a `proxy' for deliberation, where the search space is pruned not through deliberation (agents gradually converge on a single decision), but by the elimination of invalid decisions through rules. This has the added benefit of reducing the computation needed by the agents, and instead places the load on the server, thereby allowing for faster agent operation if this scenario were translated to an asynchronous system, say.



Using an optimisation paradigm favours a mathematical rule representation, whereby techniques such as gradient descent or simple estimators can be used. Therefore,  we use a matrix representation, specifying other parameters required for evaluating the rule and allowing for efficient rule retrieval. We formalise the rule representation with respect to the design considerations outlined in Section~\ref{sec:intro}, and give an example of how a declarative rule can be converted into matrix form. This gives an abstract, general-purpose rule model that can be codified for efficient computation and evaluation. 


\subsection{Rule Representation Formalism}
There are various parameters used to address the design considerations in this rule representation. Table~\ref{tab:rule_rep} illustrates this by giving the parameter name and data type used for each element in the representation. We describe the representation according to three sections: how the rules can have unique \textit{identification}, how the rules are built for efficient \textit{evaluation} and how the rules are \textit{mutable}.

\begin{table}[htb]
    \centering
    \begin{tabular}{c|c}
        \textbf{Parameter} & \textbf{Range} \\
        \hline
        ruleID & \textit{UUID} \\
        ruleName & \textit{string} \\
        ruleIsMutable & \textit{bool} \\
        ruleAction & \textit{enum} \\
        ruleInputs & [](func() $\rightarrow$ \textit{float}) \\
        ruleMatrix & [][]\textit{float} \\
        ruleComparators & []\textit{operators}
        \vspace{6pt}
    \end{tabular}
    \caption{\textit{Rule} representation data structure}
    \label{tab:rule_rep}
\end{table}

\vspace{-35pt}

\subsubsection{Identification}
Firstly, all rules have a uniquely generated \textit{ruleID} to allow all rules to be uniquely identifiable. This allows for rules to be accessed from a global lookup (a hashmap cache, say) and for agents to store a reference to the rules that they are currently using. We also supply a \textit{ruleName} for a `quality of life' benefit to the rule designer (the programmer), as this allows for a meaningful description to be added to each rule such that the programmer can see which rules are used by an agent without having to meticulously check UUIDs. The final parameter used for identification is the \textit{ruleAction}, which binds each rule to the action that they constrain. For example, a rule may need to be applied on the lootbox decision, the election of an agent to power, or the direction that the bike should travel. Binding this rule to an action allows for efficient extraction of the relevant rules, to ensure that rules which do not affect the outcome of an action (and would therefore pass as true anyway) are not evaluated, saving computation time. 

In the context of a simulator, there is likely to be a system comprising a large number of these types of rules. This could be codified with a hashmap, mapping the \textit{ruleAction} to a list of rules, for example, as this would allow for the relevant rules to be extracted in constant time.

\subsubsection{Evaluation}

There are three components used for evaluating a rule. The first is a set of \textit{ruleInputs}, or the quantitative information that the rule concerns. This allows for a rule to be suitably \textit{versatile}, as it becomes applicable to any object in the simulator, via a getter function. For example, a rule may need to be evaluated against a lootbox, concerning its position or value. A rule may also be applicable to an agent, concerning their resources (energy) or esteem. As such, specifying the constraints of a rule becomes possible with a generic getter function, allowing for a single rule engine to be applied to any kind of object (or interface, programmatically).

The second component needed for evaluating a rule is the \textit{ruleMatrix}, which is a 2-D array of numbers that applies weighting to the \textit{ruleInputs}, thereby allowing for numerical constraints to be applied. Taking the previous example of the position of a lootbox as an input variable, the \textit{ruleMatrix} may apply a weighting of 100, to define a rule that compares the (relative) position of a lootbox with a distance of 100 units. How this comparison is made is defined in the final evaluation parameter, with the \textit{ruleComparators}.

The \textit{ruleComparators} define how the input parameter is compared against its numerical weighting, using an \textit{operator} in the set $\{$ $<$, $>$, $\leq$, $\geq$, $=$ $\}$. Completing the previous example, we can define a rule that a lootbox must be within a distance of 100 units for consideration. The input parameter then becomes the lootbox position, the matrix applies a weighting of 100 and the comparator evaluates this with the 'less than or equal to' operator ($\leq$). In Section~\ref{sec:ruleRep}, we give a more complex example of a rule that concerns multiple inputs, and multiple clauses, demonstrating why a matrix is used, over a single scalar weight.

\subsubsection{Mutability}

A final design consideration is the importance of rule mutability. Given the representation as a matrix of numbers, mutability becomes trivial, as an agent/designer simply needs to change the value of a matrix element. Again, using the example of a constraint on lootbox distance, the value of 100 can be changed in the \textit{ruleMatrix} to 50, say, to give a tighter restriction on the set of feasible lootboxes. Conversely, this value may be changed to 200, to provide more `slack' on the constraint, and allow for a weaker constraint on the feasible lootboxes, and hence a wider array of possible options. Naturally, these rules may not be intended to be mutable, so we provide a \textit{ruleIsMutable} flag that dictates if the rule can be changed or not.

\subsection{Example: Lootbox Pruning} \label{sec:ruleRep}

The simplest way to define a rule in this grammar is to start with a declarative rule and convert it to a numerical representation. The (simple) rule from the previous section, \textit{A valid lootbox must be within 100 units}, can be expressed as an inequality using $d$ for relative distance as $d <= 100$.

In order to get a `better' outcome, an agent may propose that the distance should reflect the payoff of the lootbox, adding a second clause such that \textit{A valid lootbox must give a payoff of at least 1.5 times its distance}. This can also be interpreted as an inequality, with $p$ representing payoff, as $p >= 1.5*d$.







As such, we arrive at two equations that must simultaneously be true for a rule to pass. Reformatting these equations, setting them equal to zero and ascribing scalars for all variables gives:

\vspace{-1em}

\begin{align*} 
1*d + 0*p - 100*1 &<= 0 \\ 
1.5*d - 1*p + 0*1 &<= 0
\end{align*}

\noindent which can be interpreted in matrix form as:

\begin{equation}
    \begin{bmatrix}
        1 & 0 & -100 \\
        1.5 & -1 & 0
    \end{bmatrix} 
    \begin{bmatrix}
        d \\
        p \\
        1
    \end{bmatrix}
    \begin{bmatrix}
        <= \\
        <=
    \end{bmatrix}
    \overrightarrow{\textbf{0}}
\end{equation}

\noindent yielding, from left to right, the three components for rule evaluation: the \textit{ruleMatrix}, the \textit{ruleInputs} and the \textit{ruleComparators}, the result of which, after matrix multiplication, is compared against the zero vector.

\section{Evaluation of Socio-Functional Requirements}\label{sec:complexity}

In this section, we evaluate the rule representation with respect to the five \textit{socio-functional} requirements introduced in Section~\ref{sec:intro}. 

\subsection{Complexity Reduction} \label{sec:compRed}

\subsubsection{Rule Evaluation}
Given the context of a scenario with $a$ different actions, and $r$ different rules, and with repeated iteration, the importance of optimising the rule evaluation is increasingly important. Following from the rule representation, if a rule is evaluated for an irrelevant action (a lootbox rule against an election action, say) the input parameters will be unrelated to the problem (a lootbox's value isn't necessary for checking an agent's electoral eligibility), and therefore there rule will pass by default. As such, unnecessary computation is spent evaluating all rules. Given a rule matrix of size $n * m$, with $n$ clauses and $m$ input parameters, evaluation will run with $O(n*m)$ complexity. Given the full set of $r$ rules, which are evaluated for all $a$ actions, this yields the complexity of a single agent evaluating a decision as $O(n * m * r * a)$.

By ascribing a \textit{ruleAction} to the representation, as in Table~\ref{tab:rule_rep}, the rules can become \textit{stratified}, such that only a subset of the rules require evaluation. We denote this subset with $r'$. As such, by storing the active rules in a hashmap, such that the rules can be extracted in $O(1)$ time, the overall complexity is reduced to $O(n * m * r' * a)$. Given that the full ruleset is partitioned into actions, we can say that $r' * a <= r$, and as such the final complexity is $O(n * m * r)$, simplifying the complexity by a factor of $a$.

\subsubsection{Deliberation vs Social Contracts}

We can also consider a complexity improvement from the perspective of deliberation in the simulator. Previously, deliberation was the mechanism for action selection, which, in the context of \textit{Megabike} occurred in every operation phase, and therefore in every round (see Algorithm~\ref{alg:gameloop}). This meant that, for every iteration, there were at worst \textit{MaxRounds} operation phases being run. 

By transitioning to social contracts, the negotiation of rules is instead moved to the action phase, therefore being run only once per iteration. Considering $i$ iterations and $j$ rounds per iteration, with $k$ opportunities for deliberation from Section~\ref{sec:deliberation}, there were previously $O(i * j * k)$ deliberation sessions. Using social contracts allows for a single social contract negotiation session per iteration, reducing the complexity to $O(i)$.

\subsection{Linear Optimisation}

Instead of interpreting a decision as requiring the iteration of an array of distinct rules, it is possible to combine all rules into a single matrix. For example, given two distinct rules as follows (with arbitrary input parameters):

\begin{align}
    &\begin{bmatrix}
        1 & 0 & -100 \\
        3 & -1 & 0
    \end{bmatrix} 
    \begin{bmatrix}
        x \\
        y \\
        1
    \end{bmatrix}
    \begin{bmatrix}
        <= \\
        <=
    \end{bmatrix}
    \overrightarrow{\textbf{0}}
    \\
    &\begin{bmatrix}
        4 & 3 & -1 \\
        5 & -7 & 2
    \end{bmatrix} 
    \begin{bmatrix}
        z \\
        w \\
        1
    \end{bmatrix}
    \begin{bmatrix}
        > \\
        =
    \end{bmatrix}
    \overrightarrow{\textbf{0}}
\end{align}

\noindent it is possible to `stack' 
rules into a single large (potentially sparse) matrix:

\begin{equation}
    \begin{bmatrix}
        1 & 0 & 0 & 0 & -100 \\
        3 & -1 & 0 & 0 & 0 \\
        0 & 0 & 4 & 3 & -1 \\
        0 & 0 & 5 & -7 & 2
    \end{bmatrix} 
    \begin{bmatrix}
        x \\
        y \\
        z \\
        w \\
        1
    \end{bmatrix}
    \begin{bmatrix}
        <= \\
        <= \\
        > \\
        =
    \end{bmatrix}
    \overrightarrow{\textbf{0}}
\end{equation}

As such, this set of constraints may be reinterpreted as a linear optimisation problem, where agents may attempt to maximise some objective function (survivability, or energy, say) based on a set of constraints.  

\subsection{Slack: Flexibility and Mutability}

Further changes can be made to the representation to generate the ruleset as a \textit{data tableau}, so that slack variables can be used for linear optimisation. This technique synergises well with the rule representation, as the mutability of the data structure allows for not only convenient redefinition of the rule constraints but the removal or addition of extra slack for a stricter or more lenient policy, respectively. This makes the rules not only mutable, but \textit{flexible} as well.

\subsection{Enforceability and Transparency}

By having the rules bound to each \textit{Megabike}, the mutually agreed rulesets that inform all decisions are visible not only to other agents, but to the simulator designer as well. The benefit of this is twofold. Firstly, the ruleset can be seen as a tangible representation of the current agent state: that is, the algorithm they would use to decide on an action becomes publicised and interpretable in the form of a rule, i.e., instead of each agent deciding on which action to perform, which would be in the form of a black box process, they instead mutually decide on a rule which would prune the actions they wouldn't carry out. The ruleset then becomes an aggregation of all of the agents' internal processes, such that the action space yielded by evaluating the ruleset results in the set of actions that would be voted on by the agents, anyway. 


The second benefit to this representation is the ability to take computational demand away from the agents. Instead of having agents individually select an action using their own algorithm, the ruleset can be evaluated server-side instead, moving computation from individual agents onto the server. Given a synchronous system where runtime efficiency is imperative to avoiding race conditions, more complex agents can be built that don't suffer from having to run quickly to move first. Having rule evaluation performed by the server also means that the rules becomes more \textit{enforceable}; the server cannot be coerced into misevaluating a rule for personal gain, unlike if an agent were to perform this role.


\subsection{Versatility}

There are two dimensions to versatility: for the simulator and for
the simulated. Our concern here is for the simulated: what we want
to evaluate is the extent to which an agent which learns social
arrangements for the \textit{megabike scenario} could apply that
`learning' to a different scenario. This is not a requirement
that can be evaluated in the current work, but is left for future work.






\section{Empirical Simulation Results}\label{sec:simulation}


In this section, we quantitatively demonstrate the functional design considerations discussed in Section~\ref{sec:intro}: \textit{scalability}, \textit{complexity} and \textit{mutability}. For these experiments, we consider a simulator comprising 100 iterations of 100 rounds (per Algorithm~\ref{alg:gameloop}), and run the simulator 30 times to aggregate the results.

\subsection{Experiment 1: Scalability and Complexity}

\begin{figure}[htb]
    \centering
    \includegraphics[width=0.9\linewidth]{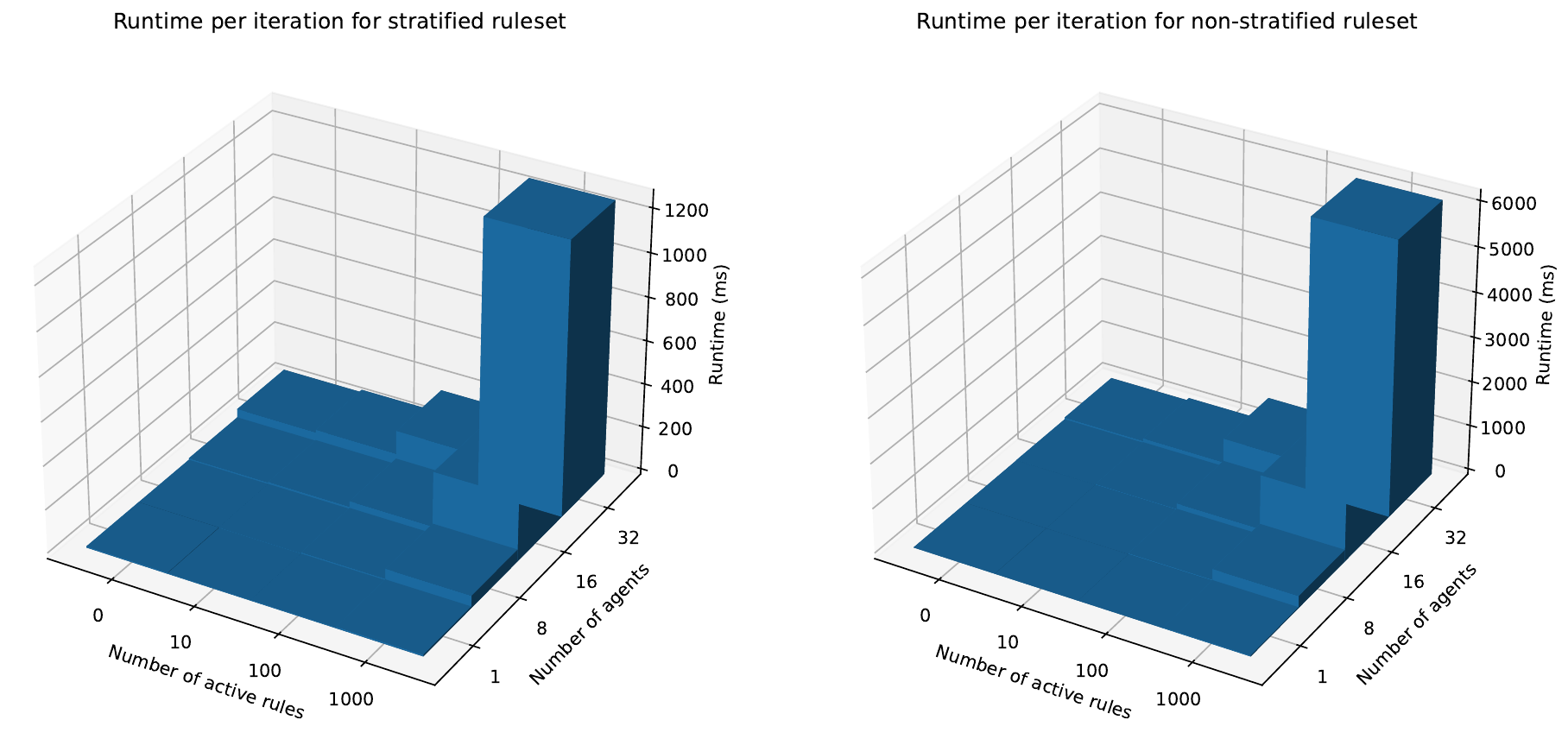}
    \caption{Runtime profiling of \textit{Megabike} simulator for stratified (left) and non-stratified (right) rulesets}
    \label{fig:runtime}
\end{figure}

This first experiment aims to demonstrate how the inclusion of rule stratification by \textit{ruleAction} results in decreased runtime for the program, and supports the claims made to reducing time complexity in Section~\ref{sec:compRed}. In this experiment, we evaluate the runtime of the program per iteration across ruleset sizes of 1, 10, 100 and 1000, and number of agents in the simulation at 1, 8, 16, and 32. The results of this experiment are shown in Figure~\ref{fig:runtime}.

To benchmark the rule representation, we need to test the worst-case runtime for the simulator. This would occur when each agent needs to evaluate every single rule. For simplicity, we define a rule that is guaranteed to pass, irrespective of the agents' state, such that the rule evaluation isn't prematurely stopped (as there is no point evaluating further rules once one has failed). To do so, we define a \textit{ruleMatrix} of all zeroes, and equate it directly with the zero vector. This rule serves as a `null' rule, which means that irrespective of the input parameters (and therefore the agent's state), the rule will pass, since all input parameters are multiplied by zero. This simplifies the calculation to whether $0 == 0$, which is always true.

The results of these experiments show that the runtime of the program per iteration is significantly reduced. Given the subset of five actions we have defined, the runtime is reduced by (approximately) a factor of five, which supports the theoretical analysis performed in Section~\ref{sec:complexity}. This shows that the rule representation is \textit{scalable} in proportion to the number of actions, agents and ruleset size. As such, more complex simulators can be built and run in a feasible time.






\subsection{Experiment 2: Mutability}

The second experiment is designed to illustrate the importance of, and simplicity in, modifying rules at runtime for survivability. In this experiment, we define a single rule that impacts the subset of lootboxes that are eligible for voting. This rule (initially) states that all eligible lootboxes must be within a radius of 1000 units. If an agent's energy falls below 50\% of the maximal capacity, agents will propose to amend the rule, once per turn, by applying a slack of 5\% (that is, increasing the radius of detection by 5\%). Alternatively, if an agent's energy is at least 50\%, the agent will propose to amend the rule by removing a slack of 5\%, thereby shrinking the radius of detection by 5\%.

\begin{figure}[htb]
    \centering
    \includegraphics[width=1.0\linewidth]{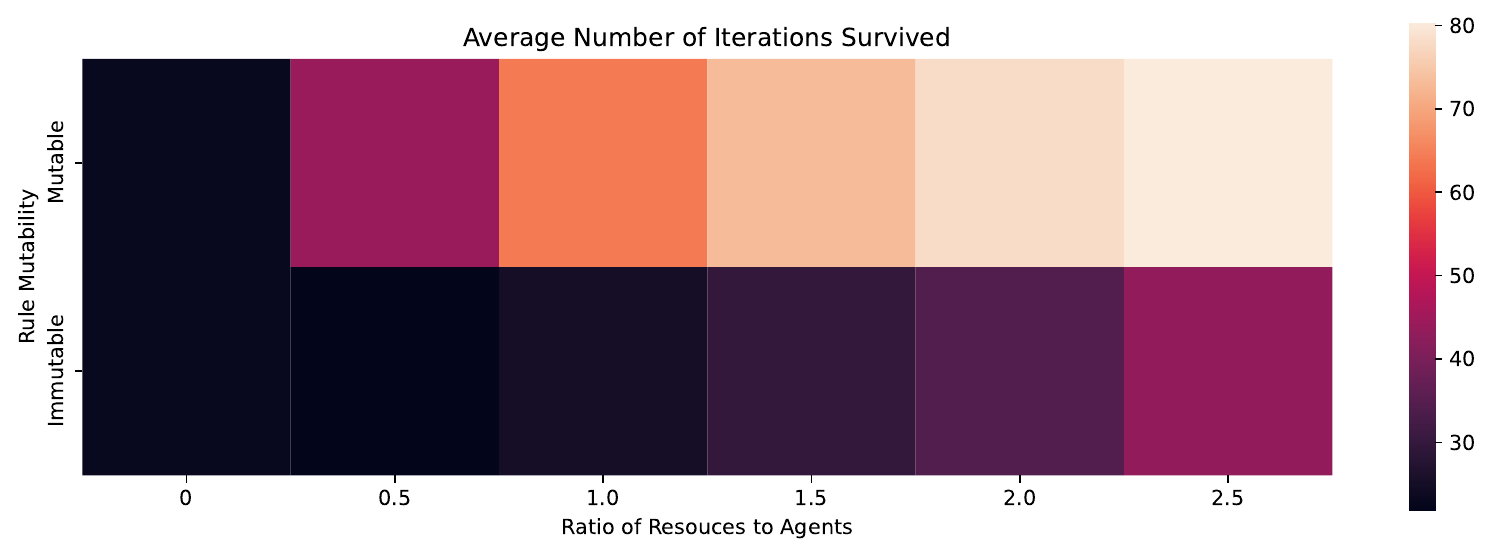}
    \caption{Average survivability of agents in \textit{Megabike} for varying degrees of resource scarcity and mutability of rules}
    \label{fig:ruleMute}
\end{figure}

This experiment varies the scarcity of resources, by varying the ratio of lootboxes to agent. Given 100 agents (across all runs), we first establish a baseline by giving a ratio of 0, thereby assessing how agents would survive given no external resources and increase this ratio to 0.5, 1.0, 1.5, 2.0 and 2.5. We also vary the capacity to mutate the rule, and illustrate the results in Figure~\ref{fig:ruleMute}.

By increasing the ratio of available lootboxes to agents, thereby alleviating the economy of scarcity, agent survivability is improved. Across all degrees of scarcity, Figure~\ref{fig:ruleMute} shows that the mutability of rules has a significant impact on survivability, outperforming the immutable rules at every stage. 

Using 0 lootboxes as a baseline, in either case, the agents survive for around 20 rounds, as the mutability of rules has no bearing on the number of lootboxes that can be achieved. In the mutable case, increasing this ratio has a drastic effect on survivability, where for an increase in 0.5x the resources, a further $\approx${20}  rounds of survival are allowed, up to $\approx${80}  when 2.5x the resources are present.

Conversely, with immutable rules, agents struggle to survive, even with 2.5x the resources, where agents still only survive for $\approx${40} rounds on average. This is due to the agents consistently depleting the nearby lootboxes over time, thereby increasing the need for a larger perception radius to detect further away resources. In the mutable case, this is exactly the kind of policy change being negotiated: the bikes consistently agree on more and more slack to be given to the rule until a sufficient number of lootboxes become visible.

\subsection{Experiment 3: Future Work in Deliberation vs Contracts}

Having demonstrated that the rule representation is appropriate for use in social contracts, we aim to further prove that these social contracts can be used as a `proxy' for social deliberation, to give a good approximation of the optimal solution. This experiment would be run over two (iterated) simulations: one where social deliberation is allowed and social contracts are not (to serve as a benchmark), and another with the opposite conditions. We would analyse these simulations by considering how `happy' the agents are with the \textit{Megabike} they are on, which is shown qualitatively by the (average) number of bike exclusions: both voluntary and involuntary.

We hypothesise that the use of social contracts should be able to approximate the number of exclusions occurring with social deliberation. For a `good' approximation, we would find that using social contracts only overshoots the number of exclusions (under deliberation) by 10\%, say, giving a confidence interval of 90\%. Furthermore, through iteration, the probability of deviating by less than 10\% should be within one standard deviation.


\section{Related Research}\label{sec:furtherWork}


Defeasibility, the property of a claim or rule to be falsified, changed, replaced, or `mutated', acts as a pillar in dynamic agent-agent communication \cite{boella2008time,panisson2014approach}. The MAS literature is not short of approaches for deciding which rules of a socio-technical multi-agent system should be changed and how. However, most approaches tackle rule change by describing processes which, despite being transparent, interpretable, and tractable, ultimately place a `heavy' burden on AI agents and MAS engineers from a socio-functional requirement standpoint.

A different perspective on addressing socio-cognitive properties is taken by the agent-oriented programming languages (AOPLs) that specify how to implement agent communication languages such as FIPA and KQML. AOPLs have previously been extended to integrate the multiple layers of abstraction in the multi-agent systems literature, which are the agent layer, the environment layer, and the organisational layer.

A modular and scalable example of the approach is the JaCaMo framework \cite{boissier2013multi}\footnote{\url{https://jacamo-lang.github.io/getting-started}}, which integrates the agent, environment and organisational layers under a single unified multi-agent-oriented programming `language'. Indeed, an important property of the rules implemented in such systems is their \textit{defeasibility}. The processes that agents of such systems follow to interact should be able to allow the agents to change the rules when it is reasonable to do so. More recently, the JaCaMo framework has been used in the MAS community for creating distributed human loop systems that leverage argumentation and mentalisation techniques, e.g., where AI agents and human users share evidence to reach more justified conclusions about each other’s mental attitudes \cite{ICAART2}.


Despite the `academic' advancements in engineering socio-cognitive MAS, fewer real-world applications of such systems have been deployed. This might be because the problem of having a speedy and cognitively efficient approach for both AI agents and experimenters regarding rule processing persists. One exception is the MAIDS framework, a JaCaMo extension, that implements intentional dialogue AI systems for human-AI teams that self-organise to optimise hospital bed allocation \cite{engelmann2023maids}. However, the hospital bed allocation problem does not have the same entangled complexity of the \textit{Megabike} scenario, which involves self-selection, self-determination of social arrangements, and existential threats. Megabike aims to provide insights into social arrangements, whereas MAIDS is tailored to a very specific domain problem.

\section{Summary and Conclusions}\label{sec:conclusions}

In summary, this workshop paper is set in the context of self-organising
multi-agent systems, in which the agents have voluntarily joined an
organisation, and now have to negotiate 
\textit{ab initio}, and repeatedly, the social arrangements for their own
self-governance.
It has specifically addressed both the abstract problem of 
balancing (ideal) social deliberation vs.\ (practical) social contracts,
and the problem of defining an expressive rule representation and efficient rule processing for these social contracts. 

The specific contributions of the current work are:
\begin{itemize}
\item to have specified the \textit{Megabike Scenario} and to define
the social arrangements designed to address the multiple
inter-dependent social problems that arise in the scenario;
\item to have identified a quintet of interleaved \textit{socio-functional}
requirements, namely scalability, complexity, mutability, 
enforceability and versatility;
\item to have discussed the contrast between social deliberation (for
which both consensus and majority decision-making can be problematic
\cite{MOP23}) and social contracts, which can be equally effective
in reaching a `correct' decision;
\item to have defined an effective representational formalism for
these social arrangements, and algorithms for efficient processing; and
\item to have derived some analytic results with respect to the socio-functional requirements, and some empirical results from simulation, that demonstrate the improved performance of social contracts
over social deliberation. 
\end{itemize}


However, the significance of this work for self-organising multi-agent systems is to highlight
that to cope with the burden of self-governance,
the agents need some awareness of their own limitations.
This includes, firstly, realising that the  
cognitive and communicative overheads imposed by deliberative decision-making
under constraints imposed by the environment is having a deleterious effect on their survivability; and secondly, 
 recognising that 
 by substituting social deliberation with social contracts they can  -- ideally -- produce approximately as good a result.
 It also demands some awareness of 
the importance of \textit{compromise} with respect to \textit{values}
in the negotiation phase of joining a \textit{megabike}, in 
deciding the original set of social arrangements, as this agreement
is the
essential assurance underpinning effective equivalence of outcomes of
social deliberation versus social contracts.

Additionally, this work suggests that there is not necessarily an `optimal'
social arrangement: there may be \textit{mutatis mutandis} more
`preferable' social arrangements according to values, but the
important requirements seem to be (a) being able to subordinate
personal preferences for benefit of the common good,
(b) being able to change `on demand' existing arrangements
to alternative arrangements that are `fit for purpose' for prevailing 
environmental conditions, and (c) not getting stuck in those arrangements when
conditions change.

\subsection*{Acknowledgements}

We are grateful for the constructive comments of three anonymous reviewers, 
which have improved this article. We also acknowledge the contribution of the
Imperial College London Department of Electrical and Electronic Engineering
SOMAS 23-24 Cohort, who developed the infrastructure used to run the experiments
described in Section~6. Thanks to Ella Bettison for assistance with that experimentation.


\bibliographystyle{splncs04}
\bibliography{references}

\end{document}